\documentclass[10pt]{article}
\usepackage{times}
\usepackage{graphicx}
\usepackage{amssymb}
\usepackage{amsmath}
\usepackage{multirow}
\usepackage[square,numbers,sort&compress]{natbib}
\newcommand{\lw}[1]{\smash{\lower1.5ex\hbox{#1}}}
\newcommand{\be}{\begin{equation}}
\newcommand{\ee}{\end{equation}\noindent}
\newcommand{\bea}{\begin{eqnarray}}
\newcommand{\eea}{\end{eqnarray}}
\newcommand{\nn}{\nonumber}

\newcommand{\maprightb}[1]{\smash{\mathop{
\hbox to 1cm{\rightarrowfill}}\limits_{#1}}}

\newcommand{\bc}{\begin{center}}
\newcommand{\ec}{\end{center}}

\newcommand{\matTwo}{\left(\begin{array}{rr}}
\newcommand{\matThree}{\left(\begin{array}{rrr}}
\newcommand{\emat}{\end{array}\right )}
\newcommand{\detTwo}{\left|\begin{array}{rr}}
\newcommand{\detThree}{\left|\begin{array}{rrr}}
\newcommand{\edet}{\end{array}\right |}

\newcommand{\Nred}{N_{\rm red}}

\begin{document}

\title{Lattice QCD and High Baryon Density State}

\author{Keitaro Nagata$^1$, Atsushi Nakamura$^1$, Shinji Motoki$^1$, \\
Yoshiyuki Nakagawa$^2$ and Takuya Saito$^3$ \\
$^1$ Research Institute for Information Science and Education, 
Hiroshima University, \\ Higashi-Hiroshima 739-8527 JAPAN \\
$^2$Graduate School of Science and Technology,  Niigata University,\\  Niigata 950-2181, Japan\\
$^3$Integrated Information Center, Kochi University, Kochi, 780-8520, Japan
}

\maketitle
We report our recent studies on the finite density QCD 
obtained from lattice QCD simulation with 
clover-improved Wilson fermions of two flavor and 
RG-improved gauge action. 
We approach the subject from two paths, i.e., the imaginary and real
chemical potentials.

\section{Introduction}

QCD at finite temperature and density has been one of the most attracting 
subjects in physics. Many phenomenological models predict that
the QCD phase diagram has a very rich structure,
and thoroughgoing analyses of heavy ion data 
show that we are sweeping finite temperature and density
regions. See Ref.~\cite{Andronic:2009gj}.

First-principle calculations based on QCD are now highly called. 
If such calculations would be at our hand,
their outcomes are also very valuable for many research fields:
high energy heavy ion collisions, the high density interior of 
neutron stars and the last stages of the star evolution.
Needless to say, the inside of nucleus is also a baryon rich
environment, and lots of contributions to nuclear physics 
could be expected.

Unfortunately, the first principle lattice QCD simulation
suffers from the sign problem.
Nevertheless, there have been many progresses such as the reweighting 
method, the imaginary chemical potential
and the canonical formulation;
now some light is  shed on the QCD phase diagram.
For reviews, see e.g.~\cite{Muroya:2003qs,deForcrand:2010ys}.

Here, we report our recent trials to promote the 
finite density lattice QCD. It contains two results \cite{Nagata:2010xi,Nagata:2011yf}:
the determination of the phase boundary of the deconfinement transition based on 
the imaginary chemical potential approach and a reduction formula for 
the Wilson fermion determinant.

\section{Imaginary Chemical Potential Approach}
\label{Ap2111sec1}

\begin{figure}[htbp]
\begin{center}
\includegraphics[height=.30\linewidth]{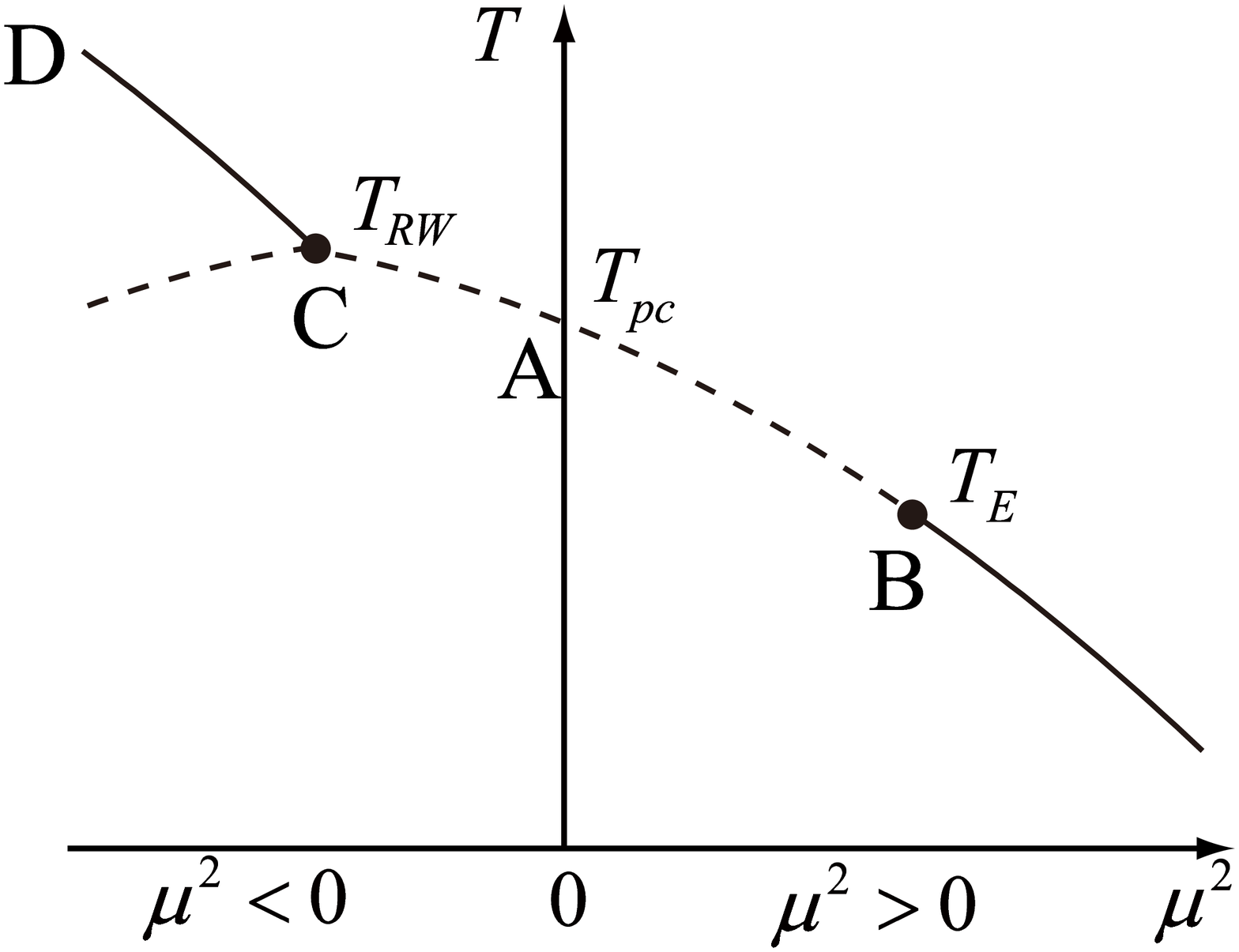}
\includegraphics[height=.30\linewidth]{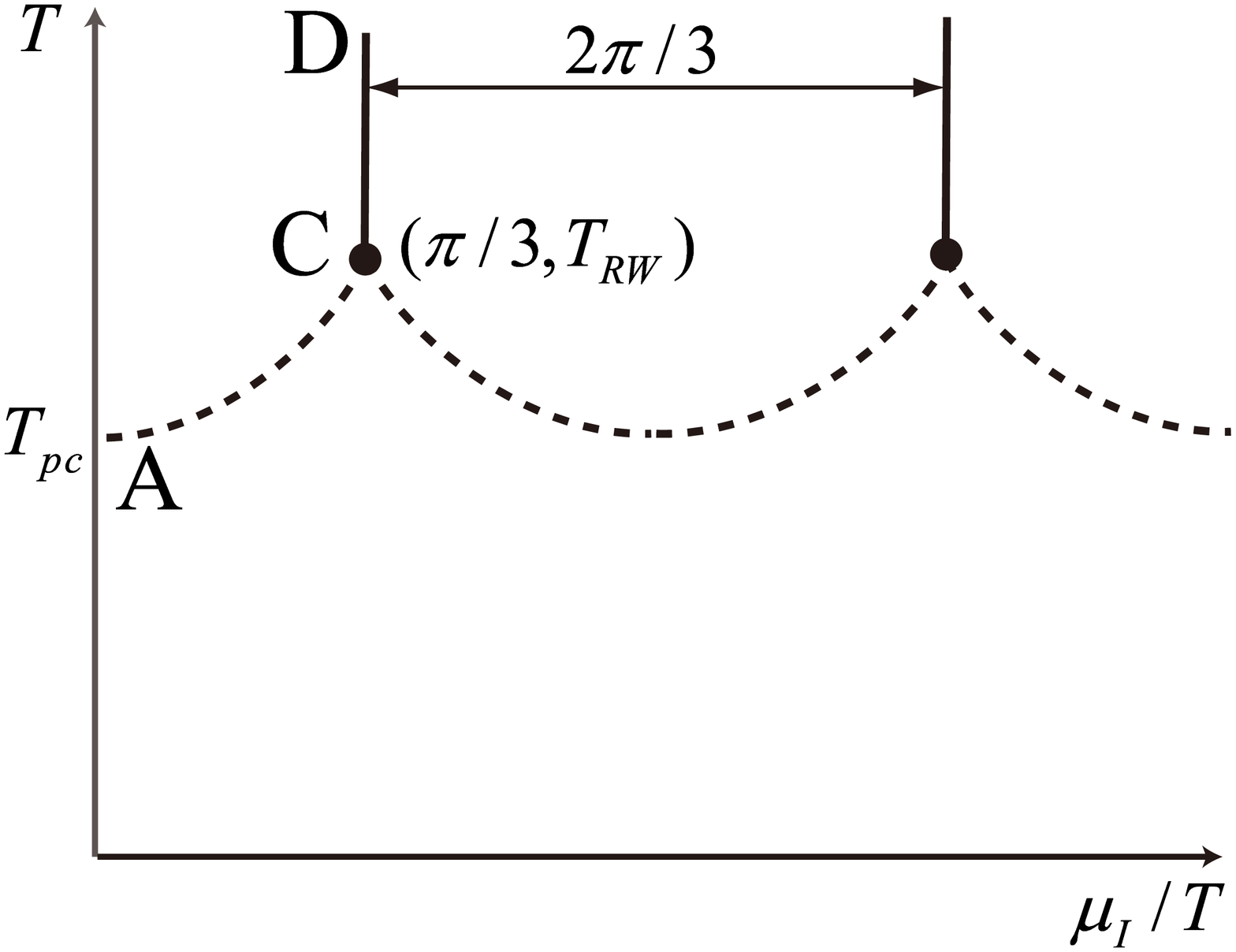}
\caption{Schematic figures for the $N_f=2$ QCD phase diagram in 
the $(\mu^2, T)$ plane (left) and $(\mu_I/T, T)$ plane (right). 
A : Pseudo-critical point at $\mu=0$. B : Critical endpoint. 
C : Roberge-Weiss endpoint. 
AB : Pseudo-critical line. AC : Extension of the line AB into the 
imaginary chemical potential plane. CD : Roberge-Weiss phase 
transition line $\mu_I/T=\pi/3$. In the right panel, larger 
$\mu_I/T$ region of the phase diagram is obtained from  the RW periodicity.
}\label{Jan2311fig1}
\end{center}
\end{figure}%
The QCD with an imaginary chemical potential is free from 
the sign problem. Using a relation
\begin{align}
(\det \Delta(\mu))^* = \det \Delta(\mu), \quad (\mu=\mu_R+i\mu_I), 
\label{Apr2411eq1}
\end{align}
it is straightforward to prove that  $\det \Delta(\mu)$ is real for $\mu=i\mu_I$. 
A partition function and its free-energy are analytic 
within one phase even if chemical potential is extended to complex, which 
is true until the occurance of a phase transition.
This validates the imaginary chemical potential approach for the study of the 
QCD phase diagram. 
In addition, the QCD phase diagram in the imaginary chemical 
potential regions have a unique feature called the Roberge-Weiss 
periodicity~\cite{Roberge:1986mm}, see Fig.~\ref{Jan2311fig1}.
There have been several studies in staggered fermions~\cite{deForcrand:2002ci,D'Elia:2009qz,D'Elia:2009tm,deForcrand:2010he,D'Elia:2002gd,D'Elia:2004at,D'Elia:2007ke,Cea:2010md,Cea:2009ba,Cea:2007vt}
and standard Wilson fermions~\cite{Wu:2006su}.

We employ a clover-improved Wilson fermion action of two-flavors and a 
renormalization-group improved gauge action. 
The clover-improved Wilson fermion action is given by 
\vspace{-5mm}
\begin{eqnarray}
&\Delta(x,y)&  =   \delta_{x, x^\prime} 
 - \kappa \sum_{i=1}^{3} \left[
(1-\gamma_i) U_i(x) \delta_{x^\prime, x+\hat{i}} 
+ (1+\gamma_i) U_i^\dagger(x^\prime) \delta_{x^\prime, x-\hat{i}}\right] \nonumber \\
 &-&
\hspace{-8mm}
\kappa \left[ e^{+\mu} (1-\gamma_4) U_4(x) \delta_{x^\prime, x+\hat{4}}
+e^{-\mu} (1+\gamma_4) U^\dagger_4(x^\prime) \delta_{x^\prime, x-\hat{4}}\right] \nonumber 
 -  \kappa  C_{SW} \delta_{x, x^\prime}  \sum_{\mu \le \nu} \sigma_{\mu\nu} 
F_{\mu\nu}.
\label{April2111eq1}
\end{eqnarray}
Here $\mu$ is the quark chemical potential in lattice unit, which is introduced to 
the temporal part of link variables.

In order to scan the phase diagram, simulations were done 
for more than 150 points on the $(\mu_I, \beta)$ plane in the domain 
$0\le \mu_I \le  0.28800$ and $1.79 \le \beta \le 2.0$. 
All the simulations were performed on a $N_s^3\times N_t = 8^3\times 4$ lattice. 
The RW phase transition line in the present setup is given by 
$\mu_I = \pi/12\sim 0.2618$. 
The value of the hopping parameter $\kappa$ were determined for each value of $\beta$ 
according to a line of the constant physics with $m_{PS}/m_V=0.8$ obtained in 
Ref.~\cite{Ejiri:2009hq}. 

\begin{figure*}[htbp]
\begin{center}
\includegraphics[height=.30\linewidth]{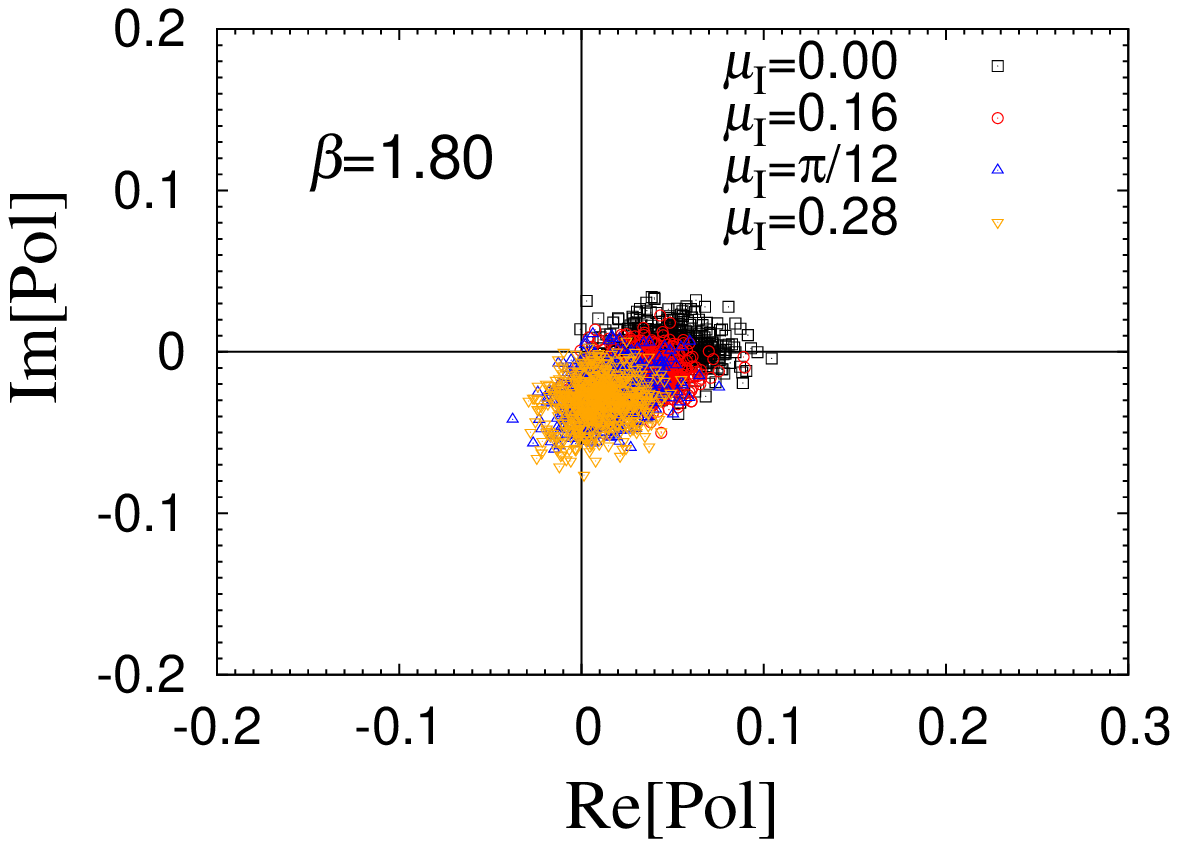}
\includegraphics[height=.30\linewidth]{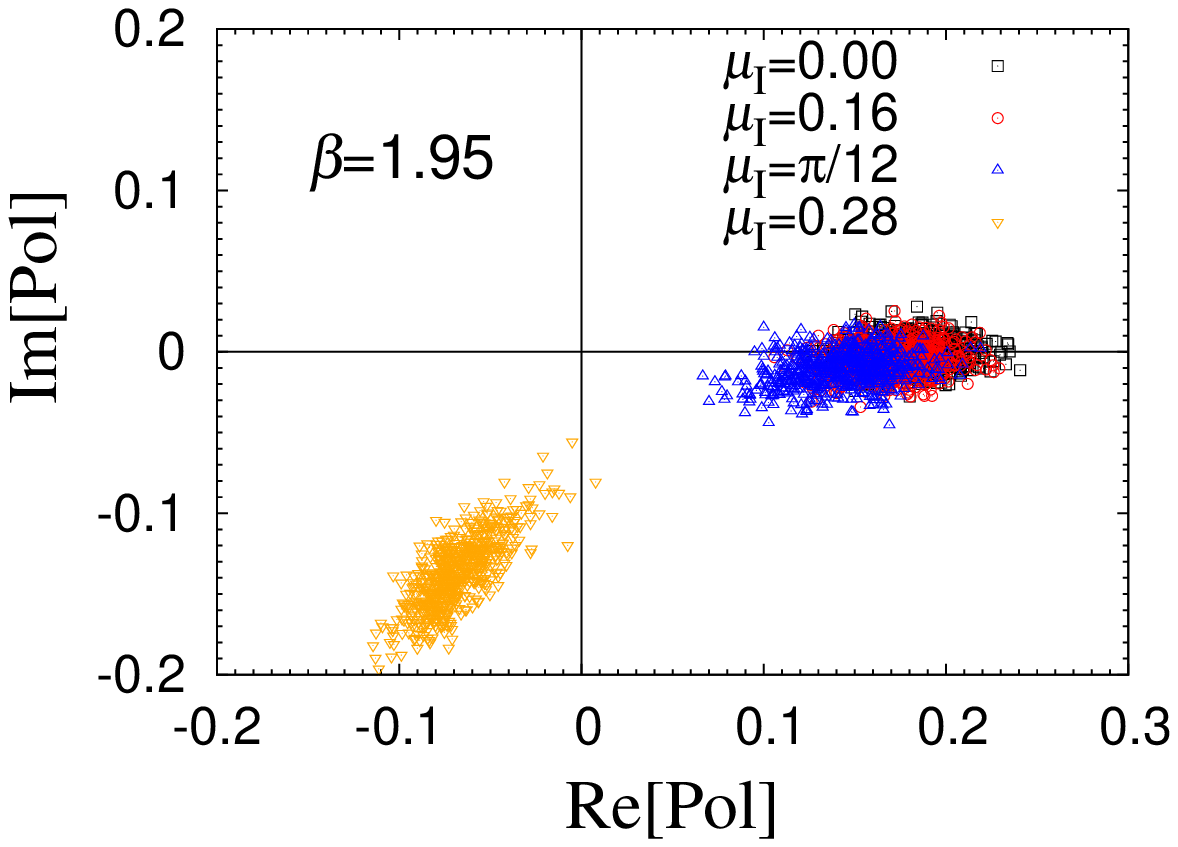}
\caption{Scatter plots of the Polyakov loop. Left : $\beta=1.80$ 
(low temperature (below $T_{pc}$)).  Right : $\beta=1.95$ (high temperature (above $T_{RW}$)). 
}\label{Mar0411fig1}
\end{center}
\end{figure*}

Scatter plots of the Polyakov loop in the complex plane  are 
shown in Fig.~\ref{Mar0411fig1}, where we choose two typical cases 
$\beta=1.80$ for the hadronic phase and $\beta=1.95$ for the QGP phase.  
At low temperatures, the Polyakov loop is small in magnitude for any
$\mu_I$ and continuously changes in a clockwise direction
as increasing $\mu_I$. 
On the other hand, at high temperatures, the Polyakov loop grows to 
$0.2\sim 0.3$. It stays at the real axis for $\mu_I<\pi/12$ and 
jumps to the left-lower side at $\mu_I=\pi/12$. 
The difference of the Polyakov loop modulus between high and low 
temperatures shows the deconfinement crossover, which is the curve 
AC in Fig.~\ref{Jan2311fig1}. 
The observed jump of the Polyakov loop at $\mu_I=\pi/12$ is 
the Roberge-Weiss phase transition, which is the line CD. 
Thus, the phase structure in $\mu^2<0$ regions of the 
QCD phase diagram can be determined by observing the behavior of 
the Polyakov loop. The properties of the phase transitions and RW endpoint 
are obtained from the susceptibility of the Polyakov loop. 
We obtain the location of  the RW endpoint $\beta= 1.927(5)$, 
which corresponds to $T/T_{pc}\sim 1.15$. 

\begin{figure*}[htbp]
\begin{center}
\includegraphics[height=.30\linewidth]{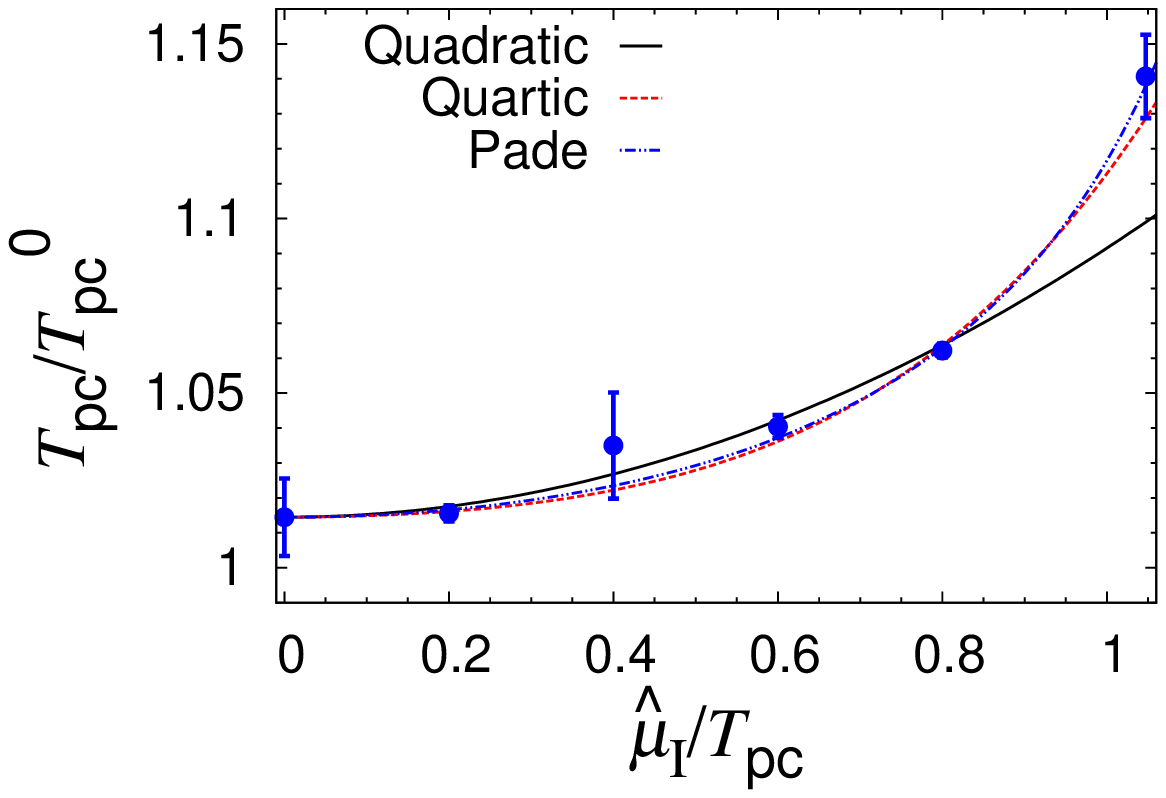}
\includegraphics[height=.30\linewidth]{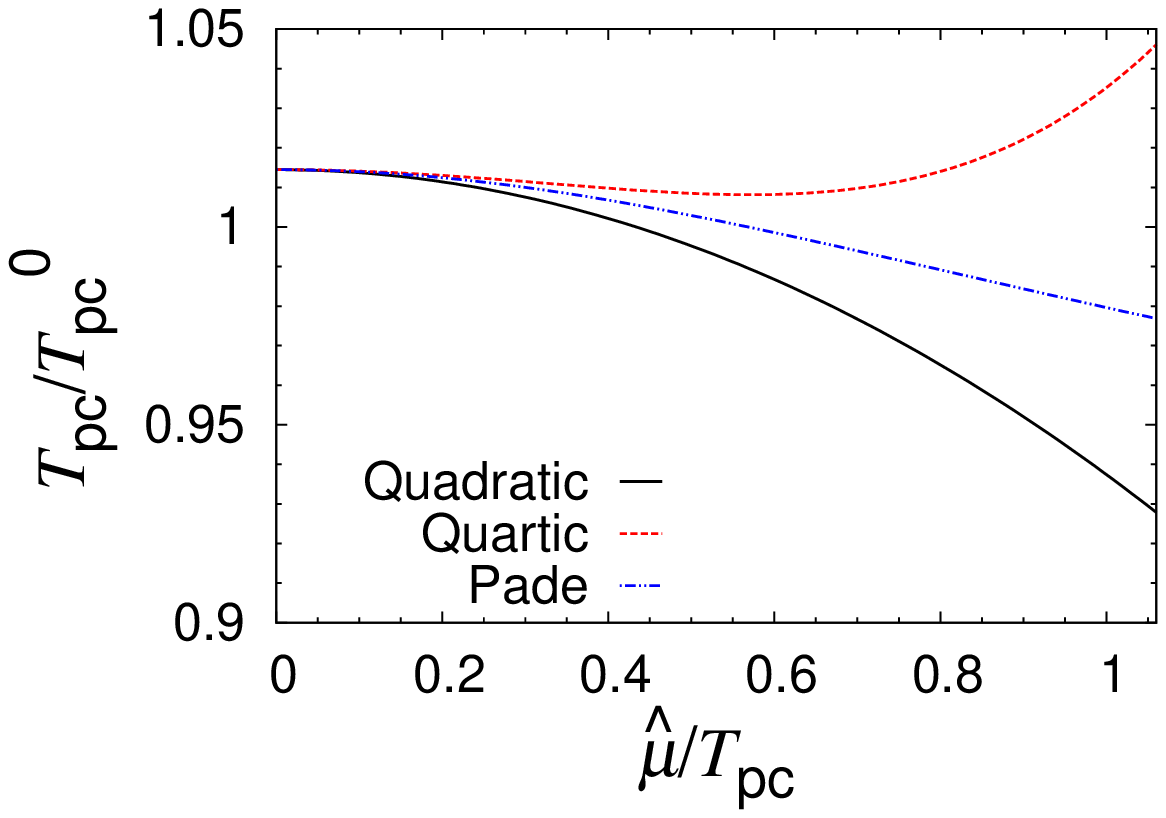}
\caption{The pseudo-critical line $\beta_{pc}$ in the imaginary 
(left panel) and real(right panel) region. 
}\label{Feb2711fig2}
\end{center}
\end{figure*}

Critical values of $\beta$ for the deconfinement transition are obtained from the 
susceptibility of the Polyakov loop modulus for each $\mu_I$. 
Using the data for the critical values of $\beta$, we can determine the pseudo-critical 
line. 
Obtained pseudo-critical line is analytically continued to $\mu^2>0$ region. 
Th results are shown in Fig.~\ref{Feb2711fig2}, where we employ physical unit 
($\mu=\hat{\mu}a$).  
The curvature at $\hat{\mu}/T_{pc}=0$ of a power series of $(\hat{\mu}/\pi T_{pc})^2$ 
is $t_2 = \pi^2 d_2 = 0.38(12)$. 
The present results are slightly smaller than other studies, 
see e.g. Ref.~\cite{Philipsen:2008gf}

\vspace{-5mm}
\section{Reduction Formula for Wilson Fermions}
\label{Apr2111sec2}

In the lattice QCD simulations with finite chemical
potential $\mu$, often we must handle the fermion determinant $\det \Delta(\mu)$, directly. 
For example, the reweighting method requires a ratio of two
determinants, $\frac{\det \Delta(\mu')}{\det \Delta(\mu)}$;
The density of state method needs the phase information\cite{Gocksch:1988iz};
The canonical formulation needs the  Fourier transformation of 
the fermion determinant.
In these approaches, the heaviest part of the numerical calculations
is the evaluation of the determinant. An efficient way of the determinant 
evaluation is highly desirable. 
Here we introduce a matrix reduction formula for Wilson fermions, which was
first constructed by Borici~\cite{Borici:2004bq}. 
Later it was studied with the inclusion of the fugacity expansion~
\cite{Nagata:2010xi,Alexandru:2010yb}. 

The Wilson fermion matrix  defined in Eq.~(\ref{April2111eq1})
can be divided into three terms 
according to their time dependence
\begin{align}
\Delta &= B -  2 z^{-1} \kappa r_- V - 2 z \kappa r_+ V^\dagger.
\end{align}
Here $r_\pm = (r \pm \gamma_4)/2$ with the Wilson parameter $r$ and $z=e^{-\mu}$. 
Each component is defined by 
\bea
B(x,x') &\equiv &   \delta_{x,x'}
 - \kappa \sum_{i=1}^{3} \left\{
        (r-\gamma_i) U_i(x) \delta_{x',x+\hat{i}}
      + (r+\gamma_i) U_i^{\dagger}(x') \delta_{x',x-\hat{i}} \right\}
+ S_{Clover}, \\
V(x,x') & \equiv & 
 U_4(x) \delta_{x',x+\hat{4}}, 
\quad
V^\dagger(x,x') \equiv
  U_4^{\dagger}(x') \delta_{x',x-\hat{4}}.
\eea
They satisfy $V  V^\dagger = I$. 
Note that $r_\pm$ are projection operators in the case that $r=1$.

Now, we construct a reduction formula for the Wilson fermions. 
A starting point is to define a permutation matrix 
$P = (c_a r_- + c_b r_+ V z^{-1})$~\cite{Borici:2004bq}.
The parameters $c_a$ and $c_b$ are arbitrary scalar except for zero, and may be set to one. 
Since $r_\pm$ are singular, the matrix $P$ must contain both of them;
otherwise $P$ is singular. It is straightforward to check 
$\det(P)  = (c_a c_b z^{-1})^{N/2} $, where 
$N=4N_c N_x N_y  N_z  N_t$. 
Multiplied by $P$, the quark matrix is transformed into
\begin{align}
\Delta P = (c_a B r_- - 2 c_b \kappa r_+ ) 
+ ( c_b B r_+ - 2 c_a \kappa r_-) V z^{-1} . 
\label{May0910eq1}
\end{align}
Carrying out the temporal part of the determinant, we obtain 
\begin{align}
\det \Delta P & = \left( \begin{array}{ccccc} 
 \alpha_1 & \beta_1 z^{-1} & & &  \\
 & \alpha_2 & \beta_2 z^{-1} & &  \\
 & & \alpha_3 & \ddots &  \\
 & & & \ddots & \beta_{N_t-1} z^{-1}\\
- \beta_{N_t} z^{-1} & & & &  \alpha_{N_t}
\end{array}\right) \nn \\
& = \left(\prod_{i = 1}^{N_t} \det(\alpha_i ) \right) 
\det\left( 1 + z^{-N_t} Q \right), 
\end{align}
where $Q = (\alpha_1^{-1} \beta_1) \cdots (\alpha_{N_t}^{-1} \beta_{N_t})$, which is often referred to as 
a reduced matrix or transfer matrix. 
The block-matrices $\alpha$ and $\beta$ are given by 
\begin{align}
\alpha_i &= \alpha^{ab, \mu\nu}(\vec{x}, \vec{y}, t_i) \nn \\
         &= c_a B^{ab, \mu\sigma}(\vec{x}, \vec{y}, t_i) \; r_{-}^{\sigma\nu} 
         -2  c_b  \kappa \; r_{+}^{\mu\nu} \delta^{ab} \delta(\vec{x}-\vec{y}), 
\\
\beta_i &= \beta^{ab,\mu\nu} (\vec{x}, \vec{y}, t_i), \nn \\ 
        &= c_b B^{ac,\mu\sigma}(\vec{x}, \vec{y}, t_i)\; r_{+}^{\sigma\nu} 
U_4^{cb}(\vec{y}, t_i) -2 c_a \kappa \; r_{-}^{\mu\nu} \delta(\vec{x}-\vec{y}) 
U_4^{ab}(\vec{y}, t_i), 
\end{align}
where the dimensions of $\alpha_i$ and $\beta_i$ are given by 
$N_{\rm red} = N/N_t = 4  N_x N_y  N_z  N_c$. 
Substituting $\det(P)  = (c_a c_b z^{-1})^{N/2} $, we obtain 
\begin{align}
\det \Delta & = (c_a c_b )^{-N/2} z^{-N/2}  \det \left( \prod_{i=1}^{N_t} \alpha_i\right)
\det\left( z^{N_t} +  Q \right).  
\label{May1010eq2}
\end{align}
Here, $Q$ is independent of $\mu$ and its rank is given by $\Nred = N/N_t$, 
while that of the Wilson fermion is originally given by $N$. 

With the eigenvalues  $\lambda_n = \{ \lambda |  \det ( Q - \lambda I) = 0 \}$, the determinant of $Q$ is given by
\be
\det(z^{N_t} + Q) = \prod_{n=1}^{N_{\rm red}} (\lambda_n + z^{N_t} ).
\label{April2011eq1}
\ee
Expanding this in powers of the fugacity $z^{N_t} = e^{-\mu/T}$,    
we finally obtain the reduced quark determinant  
\begin{align}
\det \Delta(\mu) &  =  \sum_{n=-\Nred/2}^{\Nred/2} C_n (e^{\mu/T})^n,  
\label{Jun1410eq1}
\end{align}
Note that we redefine the index of $c_n$($c_n$ by $c_{-n}$)
to obtain the second line from the first one. 
Here, $C_n = C c_n$ with 
$C = (c_a c_b )^{-N/2}\left(\prod_{i = 1}^{N_t} \det(\alpha_i ) \right)$.

\begin{figure}[htbp]
\begin{center}
\includegraphics[height=.30\linewidth]{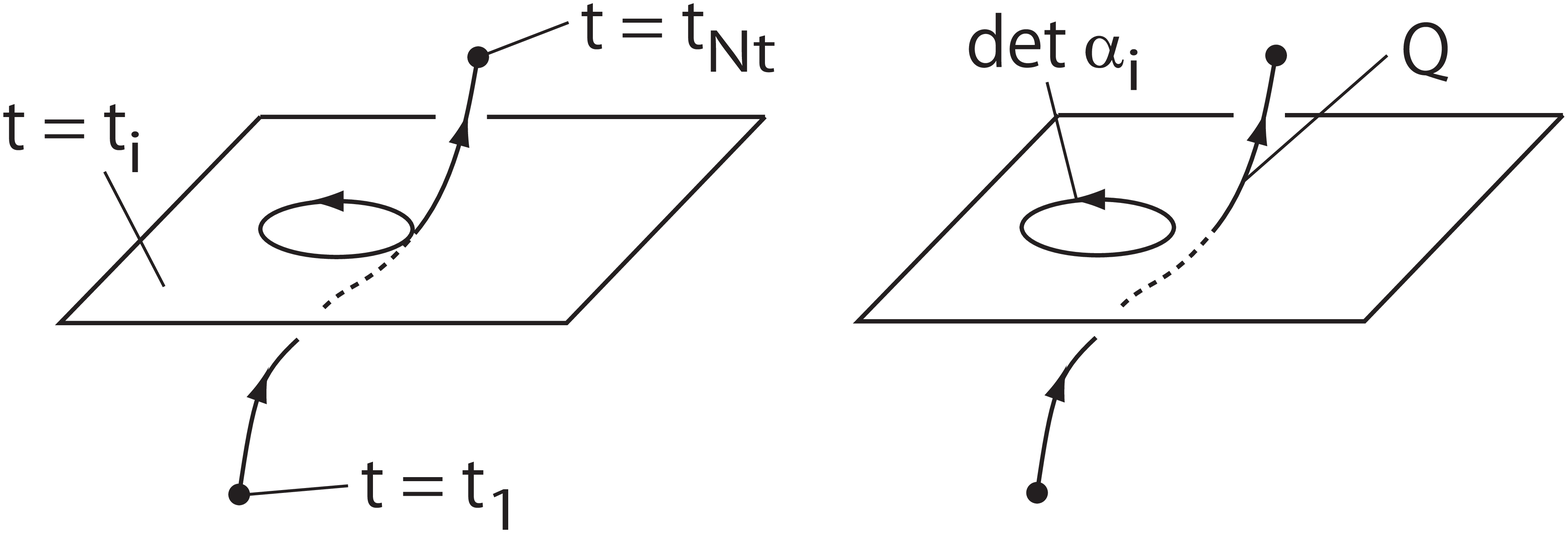}
\caption{Schematic figures for the reduction procedure.
}\label{CanJul0510fig1}
\end{center}
\end{figure}

Using a relation Eq.~(\ref{Apr2411eq1}) and the reduction formula, 
one gets
\begin{align}
 (\xi^*)^{-\frac{\Nred}{2}} \prod_{n=1}^{\Nred}(\lambda_n^* + \xi^*)
= (\xi^*)^{\frac{\Nred}{2}}\prod_{n=1}^{\Nred}(\lambda_n + (\xi^*)^{-1}), 
\label{Jan1111eq2}
\end{align}
where $\xi=z^{N_t}$. This holds for any $\xi\in \mathbb{C}$. 
For $\xi=-\lambda_n$, the left-hand side vanishes, and so should be
the right-hand side.
Then, the eigenvalue always appear in a set  
\be
\lambda_n, \quad 1/\lambda_n^*
\ee
The relation is also pointed out by 
Alexandru and Wenger ~\cite{Alexandru:2010yb}.

The reduction formula makes it easier to calculate fermion
determinant.  We plan to evaluate the phase transition line
at real chemical potential points.
Combining estimations of the phase transition line both at
real and chemical potential regions, we will get more reliable
information about QCD phase structure.

\begin{figure}
\begin{center}
  \includegraphics[height=.23\linewidth]{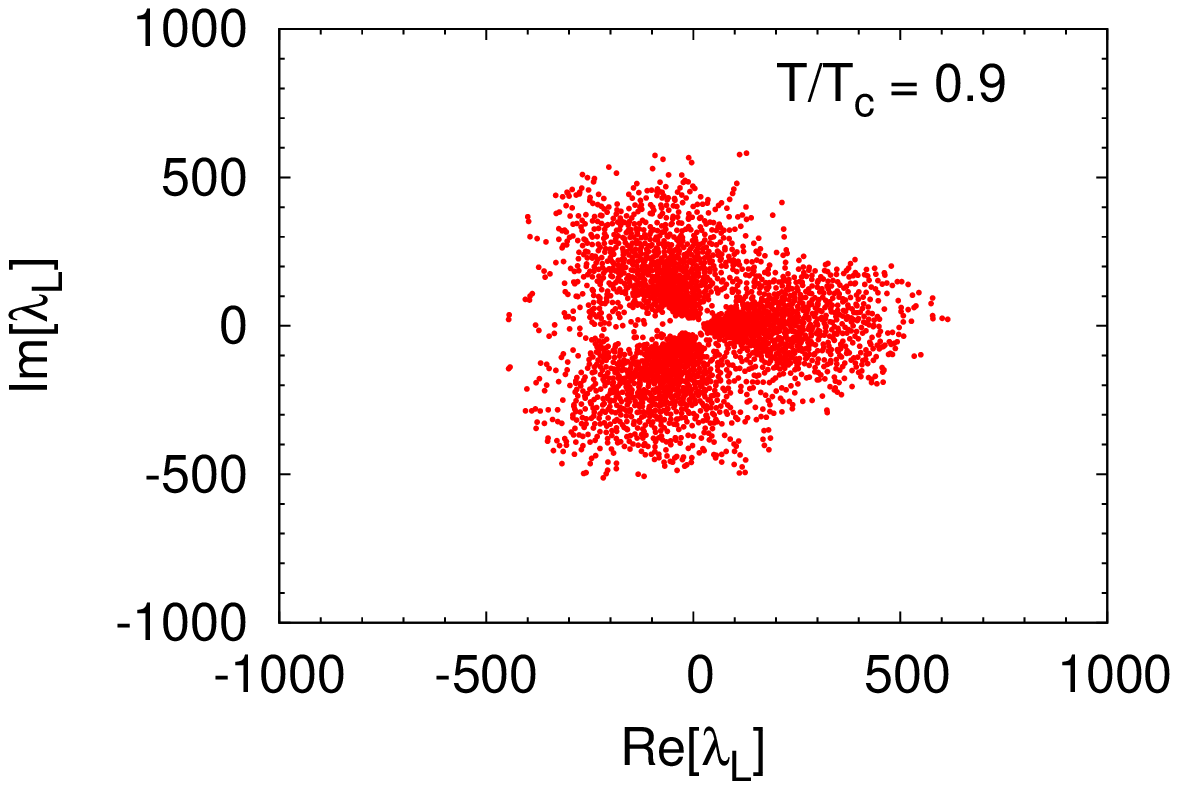}
  \includegraphics[height=.23\linewidth]{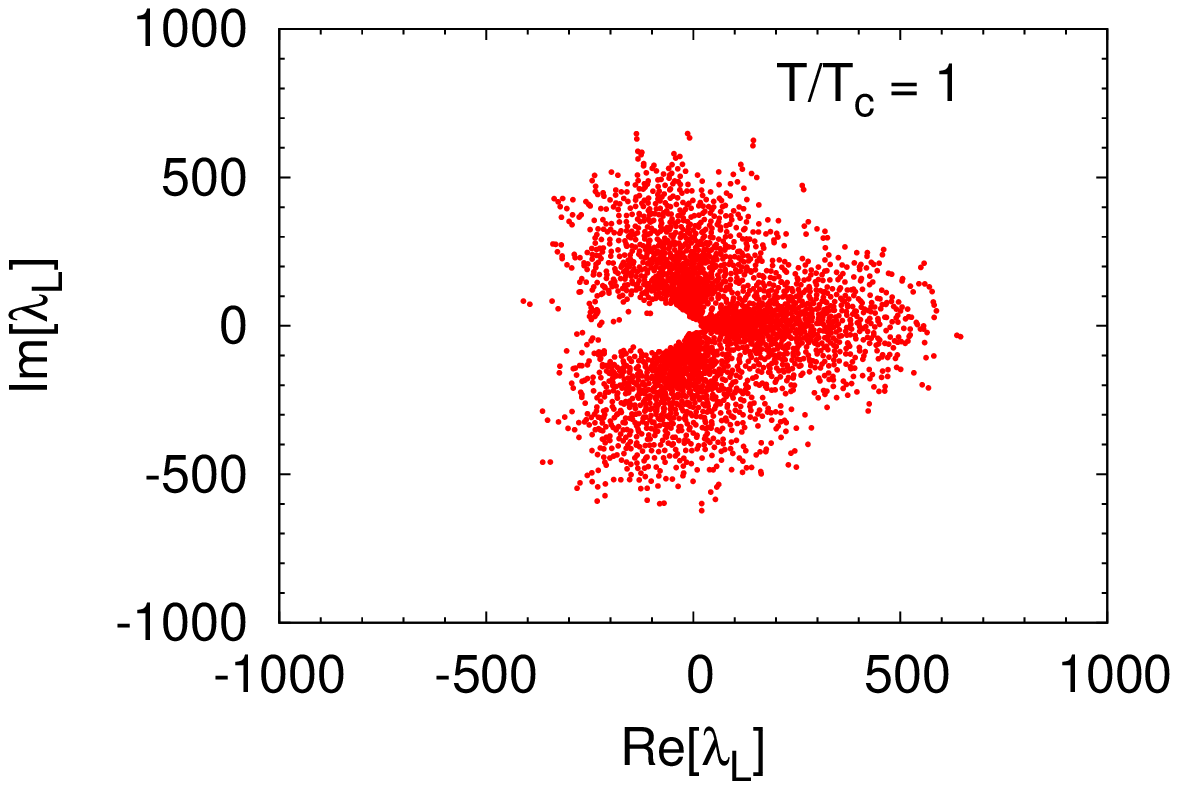}
  \includegraphics[height=.23\linewidth]{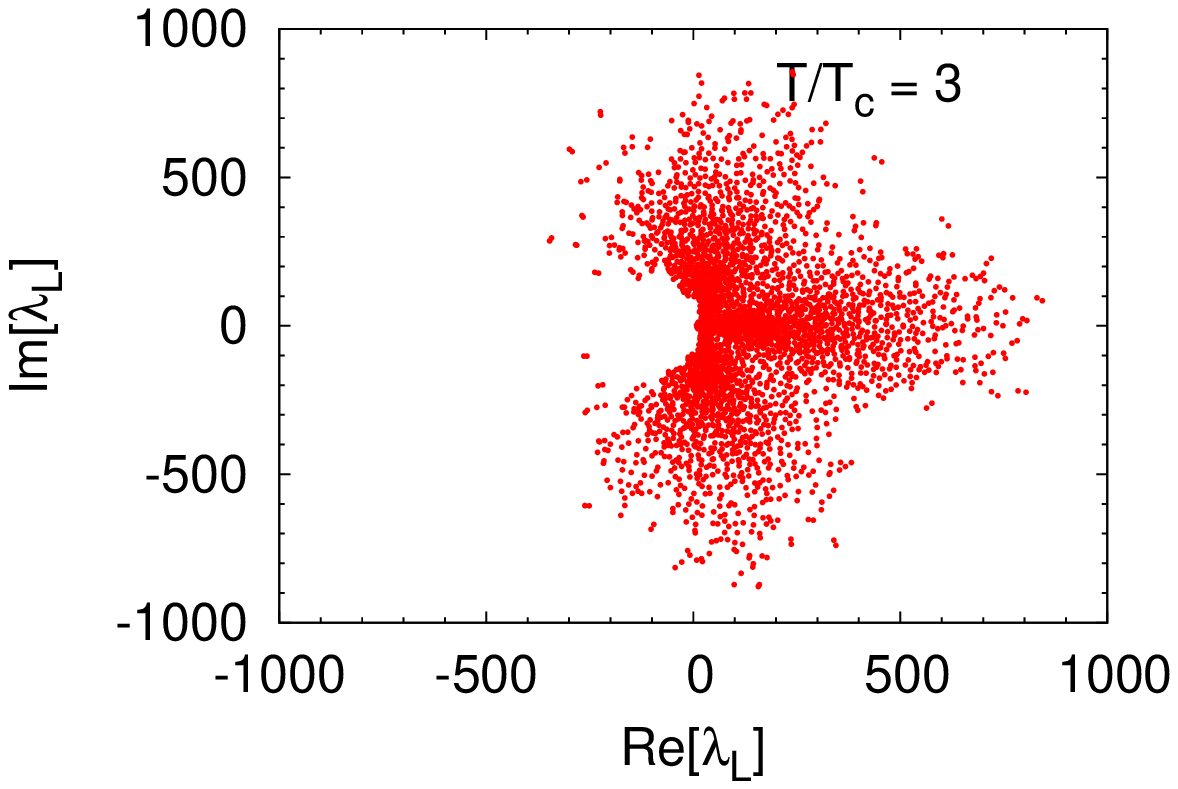}
    \caption{Distributions of the large eigenvalues in complex plane. }
    \label{Apr2011fig1}
\end{center}
\end{figure}
\begin{figure}
  \includegraphics[height=.23\linewidth]{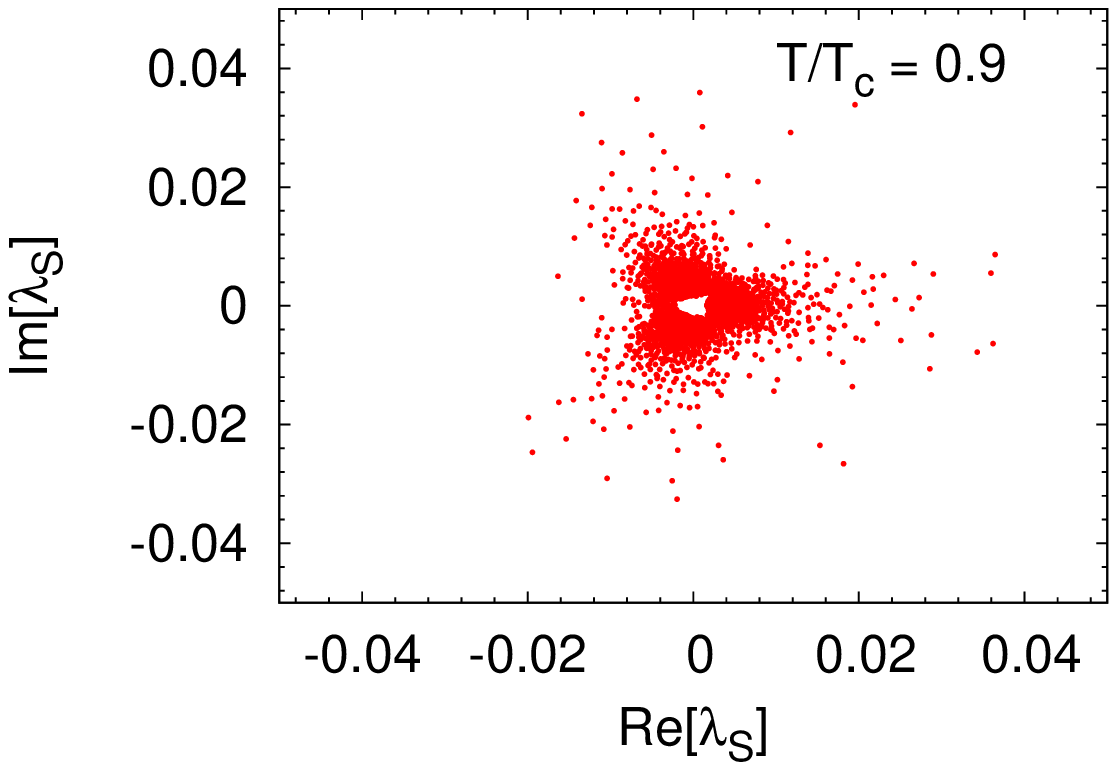}
  \includegraphics[height=.23\linewidth]{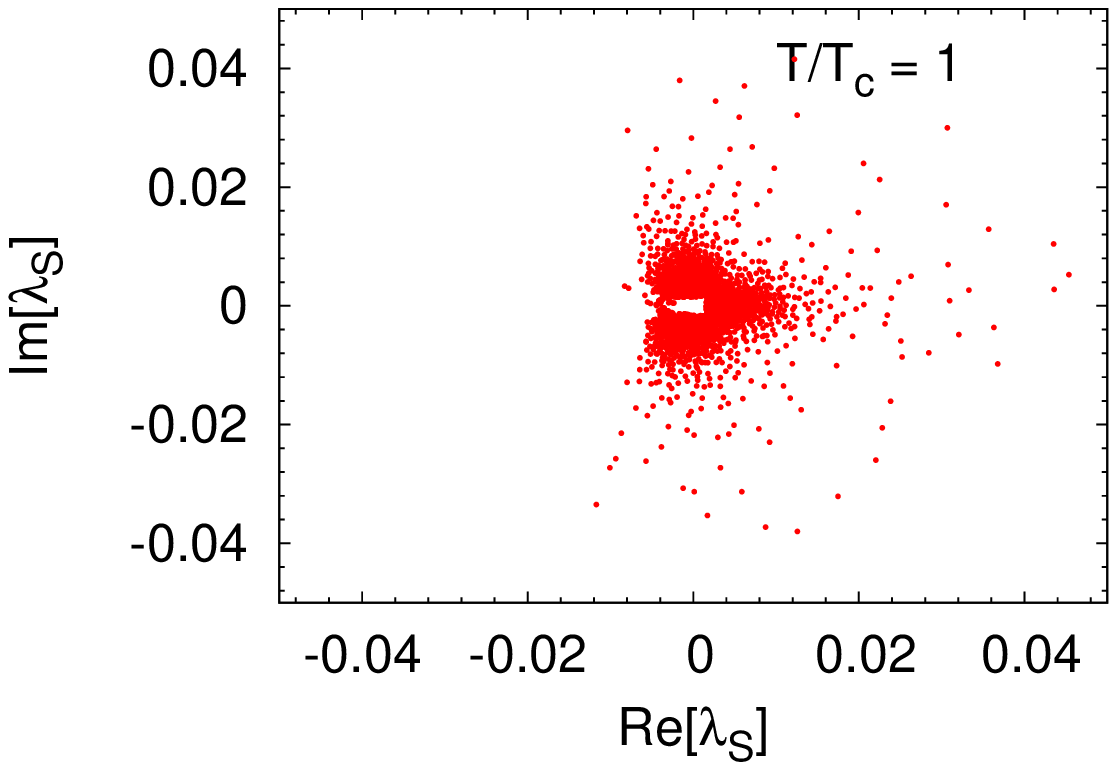}
  \includegraphics[height=.23\linewidth]{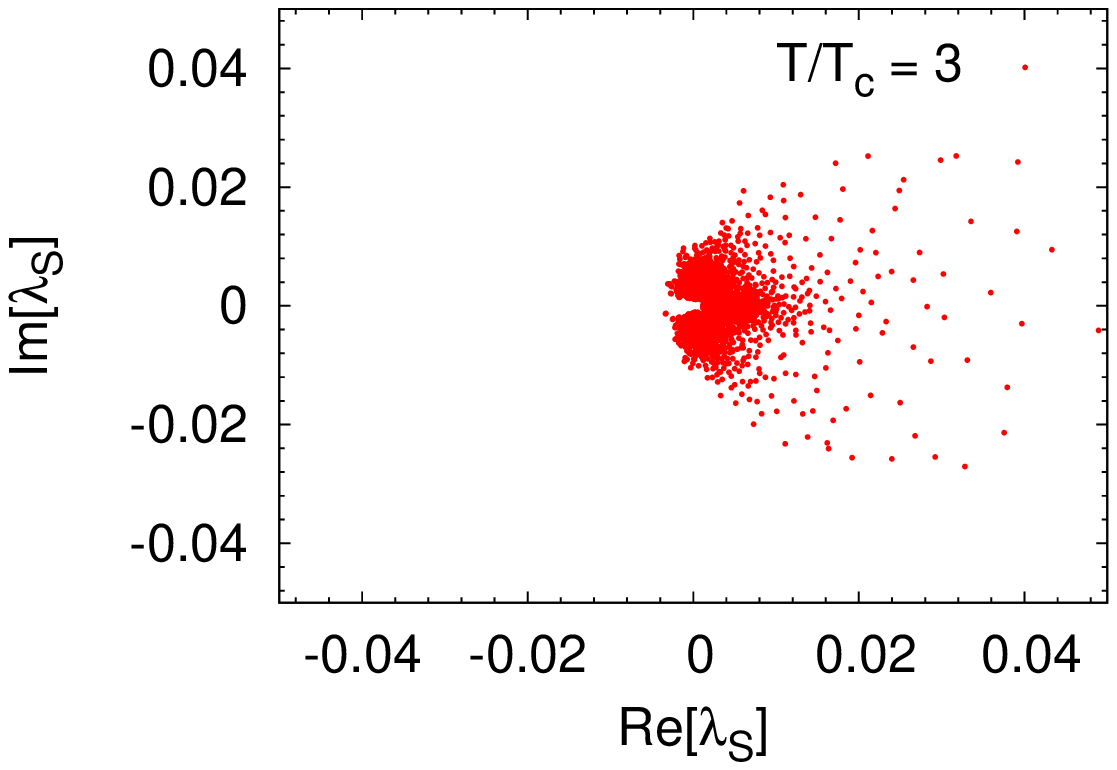}
    \caption{Distributions of the small eigenvalues in complex plane.  }
    \label{Apr2011fig2}
\end{figure}
\begin{figure}
  \includegraphics[height=.3\linewidth]{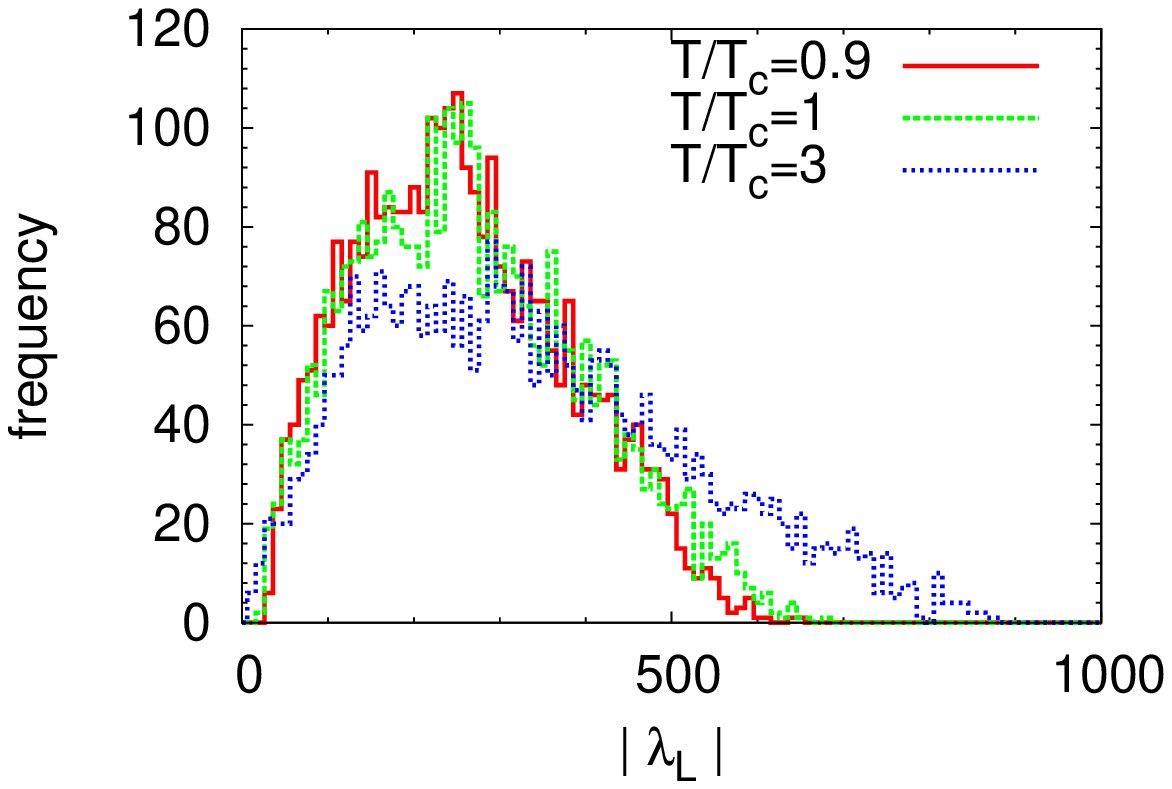}
  \includegraphics[height=.3\linewidth]{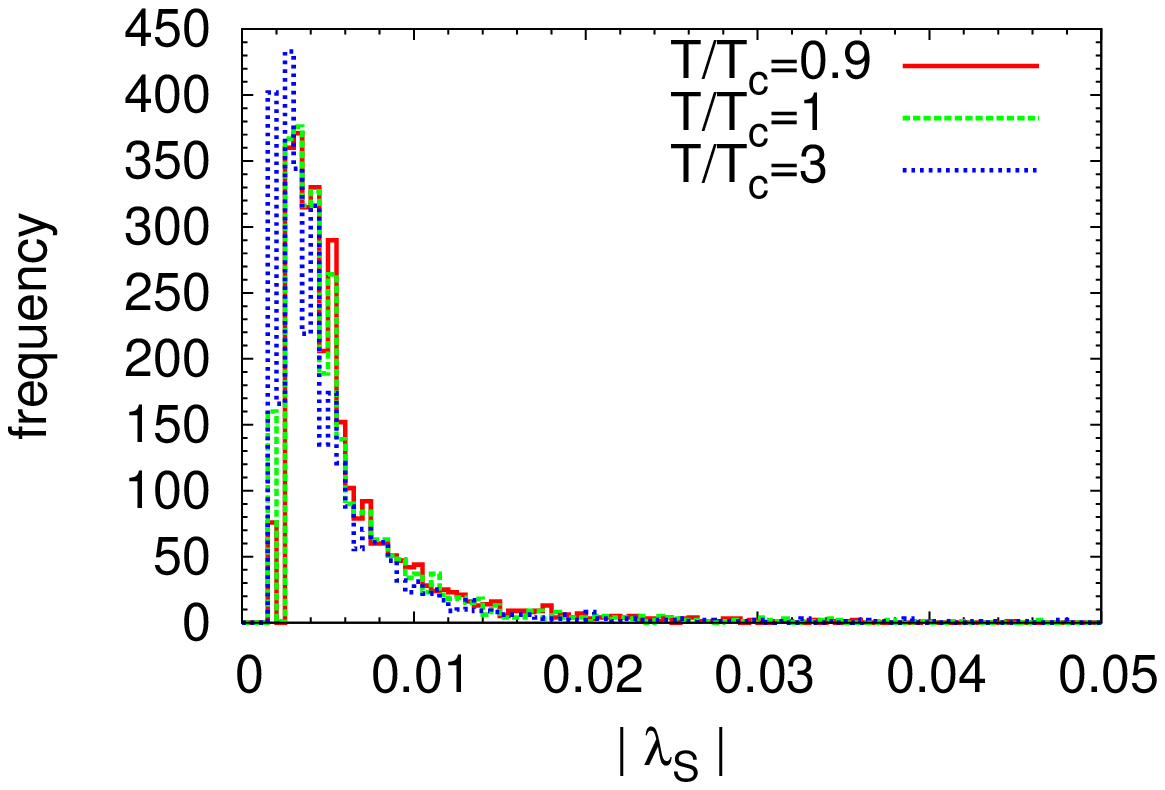}
    \caption{Histogram of the absolute value of $\lambda$. Left : large eigenvalues. Right : Small eigenvalue. }
\label{Apr2011fig3}
\end{figure}
\begin{figure}
  \includegraphics[height=.3\linewidth]{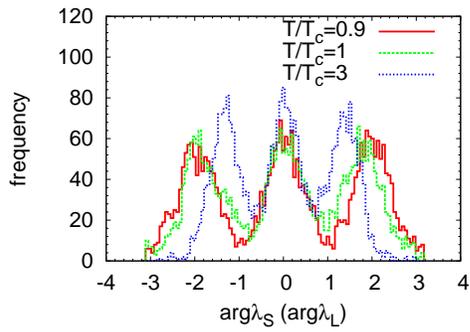}
    \caption{Histogram of the phase of $\lambda$.  }
\label{Apr2011fig4}
\end{figure}
Figures \ref{Apr2011fig1} and \ref{Apr2011fig2} show the scatter plot of 
$\{\lambda\}$  for three different temperatures 
$\beta=1.80, 1.855, 2.0$, which correspond to $T/T_c=0.9, 1, 1.3$, respectively.
Figures \ref{Apr2011fig3} and \ref{Apr2011fig4} show the histogram of the eigenvalue
distribution. 
The simulation setup was the same as given in the previous 
section. Note that Fig.~\ref{Apr2011fig2} enlarges a small domain near 
the origin in Fig.~\ref{Apr2011fig1}. 

The eigenvalues are distributed in two separate regions, and there 
is a margin between the two regions, where no eigenvalue is found. 
The histogram of the absolute value of $\lambda$ also show 
this behavior, see the right panel of Fig.~\ref{Apr2011fig3}.

The $\beta$ dependence appears in the phase of $\lambda$; 
$\lambda$ are distributed in a $Z_3$ symmetric manner at 
low temperatures, while not at high temperatures. 
The symmetric property is broken at high temperatures, and 
$\lambda$ approach to real axis. Note that this behavior is observed both 
for small and large eigenvalues, because of the pair nature of $\lambda$. 

The properties of the fugacity coefficients $c_n$ was discussed in 
\cite{Nagata:2010xi}.

The simulation was performed on NEC SX-8R at RCNP, and NEC SX-9 at CMC, 
Osaka University, and HITACHI SR11000 and IBM Blue Gene/L at KEK.
This work was supported by Grants-in-Aid for Scientific Research 20340055 and 20105003.

\baselineskip 5mm

\end{document}